\documentclass[a4paper,11pt]{article}
\usepackage{pos}

\title{Partial deconfinement in QCD
at $N=3$ and $N=\infty$}

\author*[a]{Masanori Hanada}
\author[b]{Hiroki Ohata}
\author[b]{Hidehiko Shimada}
\author[b]{and Hiromasa Watanabe}

\affiliation[a]{School of Mathematical Sciences, Queen Mary University of London\\
Mile End Road, London, E1 4NS, United Kingdom}

\affiliation[b]{Yukawa Institute for Theoretical Physics, Kyoto University\\
Kitashirakawa Oiwakecho, Sakyo-ku, Kyoto 606-8502, Japan}

\emailAdd{m.hanada@qmul.ac.uk}
\abstract{
We describe how the general mechanism of partial deconfinement applies to large-$N$ QCD and the partially-deconfined phase inevitably appears between completely-confined and completely-deconfined phases. Furthermore, we propose how the partial deconfinement can be observed in the real-world QCD with SU(3) gauge group. We propose the relationship between the behaviors of the Polyakov loop and other quantities. We test our proposal against lattice simulation data and find a nontrivial matching. }

\FullConference{The 40th International Symposium on Lattice Field Theory (Lattice 2023)\\
July 31st - August 4th, 2023\\
Fermi National Accelerator Laboratory\\}

\begin{document}
\maketitle

%%%%%%%%%%%
%%%%%%%%%%%
\section{Introduction}
%%%%%%%%%%%
%%%%%%%%%%%
Thermal phase transition in gauge theories including QCD has been important topic of research in high energy physics. 
However, the precise definition of confinement and deconfinement has been unclear in QCD, because a common definition based on the center symmetry does not hold due to the quarks in the fundamental representation and the Polyakov loop is no longer an order parameter associated with the center symmetry. Still, the Polyakov loop has been used as an ``approximate" order parameter. 

An important observation which was not well appreciated in the past is that phase transition in large-$N$ theories without center symmetry, such as QCD in the Veneziano large-$N$ limit (number of flavor $N_{\rm f}$ increases with number of colors $N$ as $N_{\rm f}/N$ is fixed) can be described by the Polyakov loop. Furthermore, even in theories with center symmetry such as pure Yang-Mills, there is a phase transition between two center-broken phases which is often called Gross-Witten-Wadia (GWW) transition. 
These facts suggested that Polyakov loop has a meaning unrelated to center symmetry. 

It turned out that Polyakov loop is related to the only ``symmetry" shared by all examples discussed in the past: gauge symmetry~\cite{Hanada:2020uvt}. Specifically, each quantum state in the operator formalism has different amount of gauge redundancy captured by the Polyakov loop. Confined vacuum has no redundancy, while deconfined states have large redundancy that increases with $N$ as $e^{N^2}$. The amount of redundancy has important consequence that can easily be understood by noticing a simple but deep connection between color confinement and Bose-Einstein condensation~\cite{Hanada:2020uvt}. 

In this conference proceeding, we discuss how these findings can be generalized to finite-$N$ theories, following Refs.~\cite{Hanada:2023rlk,Hanada:2023krw} 
In Sec.~\ref{sec:large-N}, we review the findings for large-$N$ theories. We will also give some new results. 
In Sec.~\ref{sec:finite-N}, we discuss real-world QCD. Sec.~\ref{sec:conclusion} is for conclusion and discussion. 
%%%%%%%%%%%%%%%%%%%%
%%%%%%%%%%%%%%%%%%%%
\section{Thermal transition at $N=\infty$}\label{sec:large-N}
%%%%%%%%%%%%%%%%%%%%
%%%%%%%%%%%%%%%%%%%%
Let us consider large-$N$ QCD in the Veneziano limit with sufficiently light quark mass mimicking real-world QCD. It is not hard to imagine that we can avoid a first order transition. Then, do we expect a crossover rather than a phase transition? In fact, with mild assumptions, we can say there must be a phase transition. 

Let $\theta$ and $\rho(\theta)$ be a phase of Polyakov line and its distribution, respectively. 
The distribution $\rho(\theta)$ is defined at $-\pi\le\theta<\pi$ and normalized as $\int_{-\pi}^\pi {\rm d}\theta\rho(\theta)=1$.  
At zero temperature, the distribution is constant: $\rho(\theta)=\frac{1}{2\pi}$. 
At infinitely high temperature, it is the delta function: $\rho(\theta)=\delta(\theta)$. 
Suppose that \textit{the density changes continuously}. Then, \textit{there must be a point where a gap is formed in  $\rho(\theta)$.} The opening of a gap is typically a phase transition in the large-$N$ limit, such as the Gross-Witten-Wadia (GWW) transition. Below, we use the word `GWW transition' to denote the transition associated with the formation of a gap. 

\begin{figure}[htbp]
\begin{center}
\scalebox{0.17}{
\includegraphics{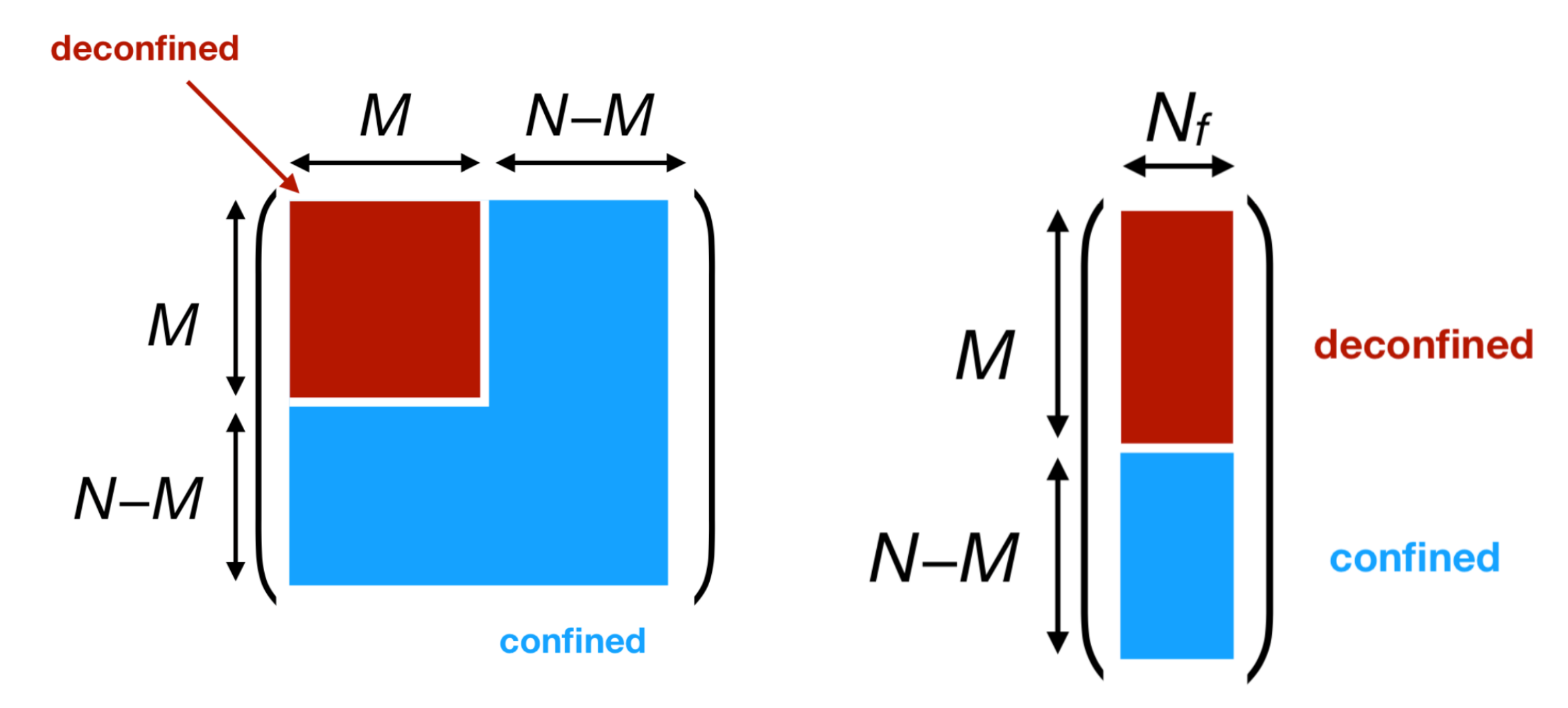}}
\end{center}
\caption{
Partial deconfinement in SU($N$) QCD with $N_f$ fundamental quarks. [Left] Gauge field. $M\times M$ sub-matrix is deconfined.
[Right] Quarks. $M$ components are deconfined. Any embeddings of SU($M$) into SU($N$) are equivalent because of gauge symmetry. This figure is taken from Ref.~\cite{Hanada:2019czd}.  
}\label{fig:partial_deconfinement}
\end{figure}

The GWW transition separates completely-deconfined phase and partially-deconfined phase~\cite{Hanada:2020uvt,Hanada:2018zxn,Hanada:2016pwv,Berenstein:2018lrm}. To understand it, we write thermal partition function in two different but equivalent ways: 
\begin{eqnarray}
    Z(T) = \mathrm{Tr}_{\mathcal{H}_{\text{inv}}} e^{-\hat{H}/T} = \frac{1}{\text{vol}\mathcal{G}}\int_\mathcal{G} dg \mathrm{Tr}_{\mathcal{H}_{\text{ext}}}\left(\hat{g} e^{-\hat{H}/T} \right)
\end{eqnarray}
where $\mathcal{G}$ is the set of all gauge transformations (for the SU($N$) gauge theory on a lattice, $\prod_{\vec{x}}[\mathrm{SU}(N)]_{\vec{x}}$ where $\vec{x}$ runs for all spatial lattice points), $\mathcal{H}_{\text{ext}}$ is the extended Hilbert space that contains nonsinglet states, $\mathcal{H}_{\text{inv}}$ is the gauge invariant Hilbert space, and the integral of $g\in\mathcal{G}$ is taken over the Haar measure. A simple but crucial fact is that $\hat{g}$ is the Polyakov line~\cite{Hanada:2020uvt}. An energy eigenstate $|\Phi\rangle\in\mathcal{H}_{\text{ext}}$ contributes to partition function as
\begin{eqnarray}
    \frac{1}{\text{vol}\mathcal{G}}\int_\mathcal{G} dg \langle\Phi|\hat{g}|\Phi\rangle e^{-E_\Phi/T}\, . 
\end{eqnarray}
Obviously, if $\int_\mathcal{G} dg \langle\Phi|\hat{g}|\Phi\rangle$ is large, contribution from such a state is enhanced. It turns out that confined vaccuum has a large enhancement factor, while deconfined states do not. More generally, SU($M$)-deconfined states in the partially-deconfined phase (Fig.~\ref{fig:partial_deconfinement}) has enhancement factor that scales as $e^{(N-M)^2\times\mathrm{volume}}$. 

Typical Polyakov lines dominating the path integral are the ones that contribute to the enhancement factor $\int_\mathcal{G} dg \langle\Phi|\hat{g}|\Phi\rangle$ for a typical state $|\Phi\rangle$. For the confined vacuum, typical Polyakov lines are slowly varying Haar random~\cite{Hanada:2023rlk}, as explained below. If we look at one-point function, we simply see the Haar random distribution. Note that the constant distribution $\rho(\theta)=\frac{1}{2\pi}$ is the Haar random distribution of SU($\infty$). Furthermore, SU($N$) Haar randomness is a stronger condition than $\mathbb{Z}_N$ center symmetry; this is the reason that expectation value of the Polyakov loop vanishes at low temperature even in QCD. The Polyakov loop has a meaning not related to center symmetry!

In the extended-Hilbert space picture, the confined ground state in or sufficiently close to the continuum limit is a wave packet localized around $U_{\vec{n},\mu} = \mathbf{1}_N$ up to gauge transformation. Any wave packets connected to this wave packet by a gauge transformation $\mathbf{1}_N \rightarrow \Omega_{\vec{n}}^{-1} \Omega_{\vec{n}+\mu}$ ($\Omega \in SU(N)$) are equivalent. 
Let us consider the action of Polyakov loop $P_{\vec{n}} \equiv \Omega_{\vec{n}}^{-1} V_{\vec{n}} \Omega_{\vec{n}}$ on such a vacuum configuration. (Or, we could fix the gauge such that the ground state is a wave packet around $\mathbf{1}_N$ and drop $\Omega$ from the discussion below.)  
It is straightforward to see that 
\begin{eqnarray}
    P_{\vec{n}}^{-1} (\Omega_{\vec{n}}^{-1} \Omega_{\vec{n}+\hat{\mu}}) P_{\vec{n}+\hat{\mu}} = \Omega_{\vec{n}}^{-1} V_{\vec{n}}^{-1} \Omega_{\vec{n}} (\Omega_{\vec{n}}^{-1} \Omega_{\vec{n}+\hat{\mu}}) \Omega_{\vec{n}+\hat{\mu}}^{-1} V_{\vec{n}+\hat{\mu}} \Omega_{\vec{n}+\hat{\mu}} = \Omega_{\vec{n}}^{-1}(V_{\vec{n}}^{-1} V_{\vec{n}+\hat{\mu}})\Omega_{\vec{n}+\hat{\mu}}\, . 
\end{eqnarray}
Therefore, the vacuum is invariant if $V_{\vec{n}}$ is constant, and approximately invariant if $V_{\vec{n}}$ is slowly varying. There is no constraint otherwise, so the statistical distribution is Haar random. Such an approximate invariance under the slowly varying Haar random transformation leads to an enhancement as in the BEC case, but with much larger enhancement factor $\sim e^{N^2}$. 

For partially-deconfined states shown in Fig.~\ref{fig:partial_deconfinement}, 
$
V_{\vec{n}} = 
\left(
\begin{array}{cc}
\textbf{1}_{M} & \textbf{0}\\
\textbf{0} & \tilde{V}_{\vec{n}}
\end{array}
\right)
$
with slowly varying $\tilde{V}_{\vec{n}}\in\textrm{SU}(N-M)$, leads to the enhancement factor $\sim e^{(N-M)^2}$. Because of this enhancement factor, at intermediate energy scale, the deconfined sector curls up to the SU($M$)-subgroup of SU($N$). 
%%%%%%%%%%%%%%%%%%%
%%%%%%%%%%%%%%%%%%%
\subsubsection*{Continuum limit and renormalization}
%%%%%%%%%%%%%%%%%%%
%%%%%%%%%%%%%%%%%%%
The discussion above used bare Polyakov loop. The symmetry of wave function on each spatial link is captured by bare Polyakov loop, but if lattice spacing is sent to zero, symmetry at the UV cutoff scale does not necessarily capture the low-energy physics we are interested. Indeed, in the continuum limit, the expectation value of the bare Polyakov loop in any nontrivial representation vanishes even in the deconfined phase and the bare Polyakov line becomes Haar random. Through the character expansion, Haar randomness of the Polyakov line follows. 
From the symmetry point of view, this is because the wave function on each link cannot be distinguished from the ground state when we zoom in to ultraviolet. Still, nontrivial low-energy properties at low energy appears because the number of links grows. 
Properly renormalized~\cite{Gupta:2007ax,Mykkanen:2012ri} or smeared~\cite{Petreczky:2015yta,Datta:2015bzm} Polyakov line captures sensible low-energy physics.

In the current study, we do not expect that the renormalization affects the analyses because the configuration set we use has only one lattice spacing and temperature is varied by changing the lattice size.

%%%%%%%%%%%%%%%%%%%
%%%%%%%%%%%%%%%%%%%
\section{Thermal transition at $N=3$}\label{sec:finite-N}
%%%%%%%%%%%%%%%%%%%
%%%%%%%%%%%%%%%%%%%
The notion of the size of deconfinement sector $M$ could be subtle for $N=3$, because the $1/N$ correction could be as large as $M$ itself. Still, $M=0$ does not have ambiguity. The confined vacuum is invariant under $\mathrm{SU}(N)$, and the Polyakov lines are Haar-random distributed. (Here we consider only the one-point functions.) In particular, the distribution of the phases should be
\begin{align}
\rho_{\rm Haar}(\theta)
=
\frac{1}{2\pi}
\left(
1
-
(-1)^N\cdot \frac{2}{N}\cos(N\theta)
\right)\, . 
\end{align}

Complete confinement is characterized by slowly varying Haar random distribution of Polyakov line up to small corrections. 
Specifically, expectation value of a Polyakov loop in an irreducible representation r is expected to behave as $e^{-m_{\rm r}/T}$, where $m_{\rm r}\ge 0$ is the mass of the lightest excitation in this representation. Note that exact Haar randomness means vanishing expectation values of all loops in any representation r. For QCD, because of the mass gap, all $m_{\rm r}$s should be positive and the corrections to Haar randomness should be parametrically small at low temperature, possibly consistent with zero in our resolution.  

\begin{figure}[htbp]
\begin{center}
\scalebox{0.2}{
\includegraphics{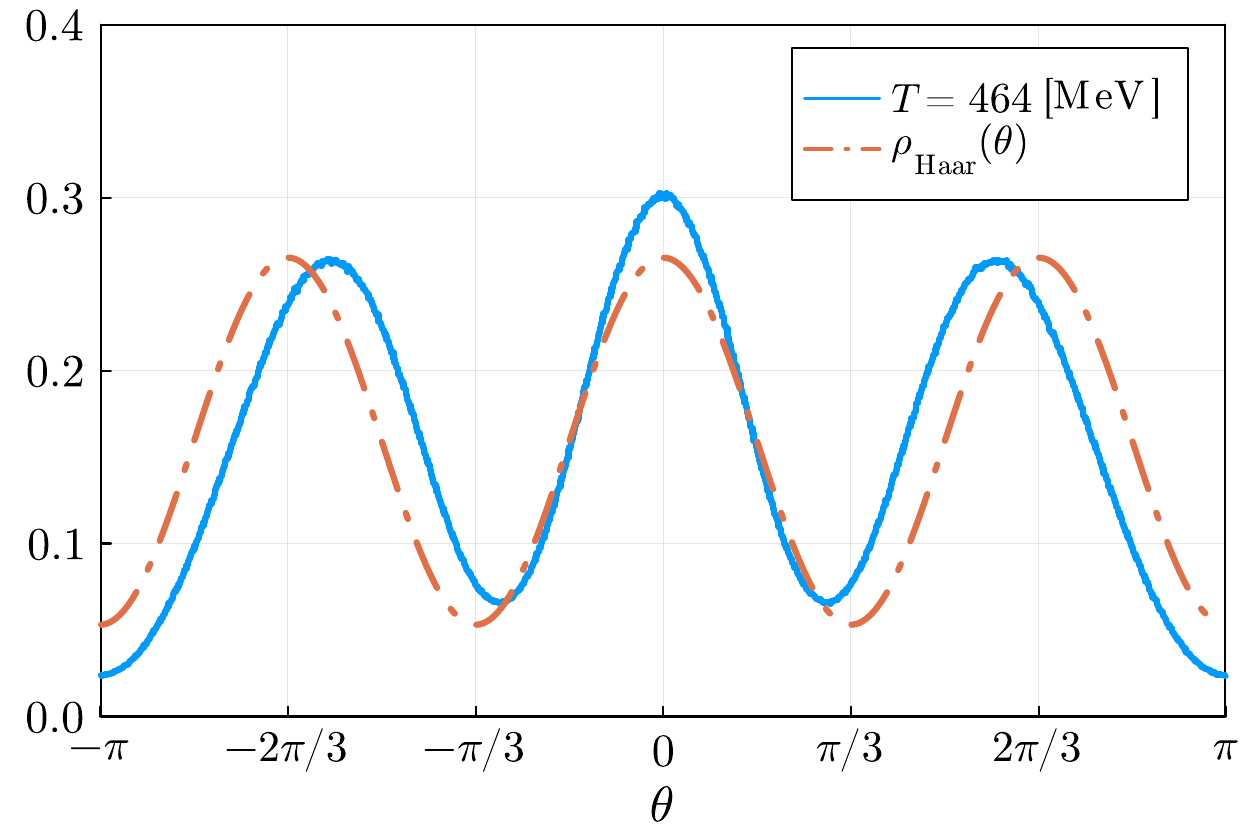}}
\scalebox{0.2}{
\includegraphics{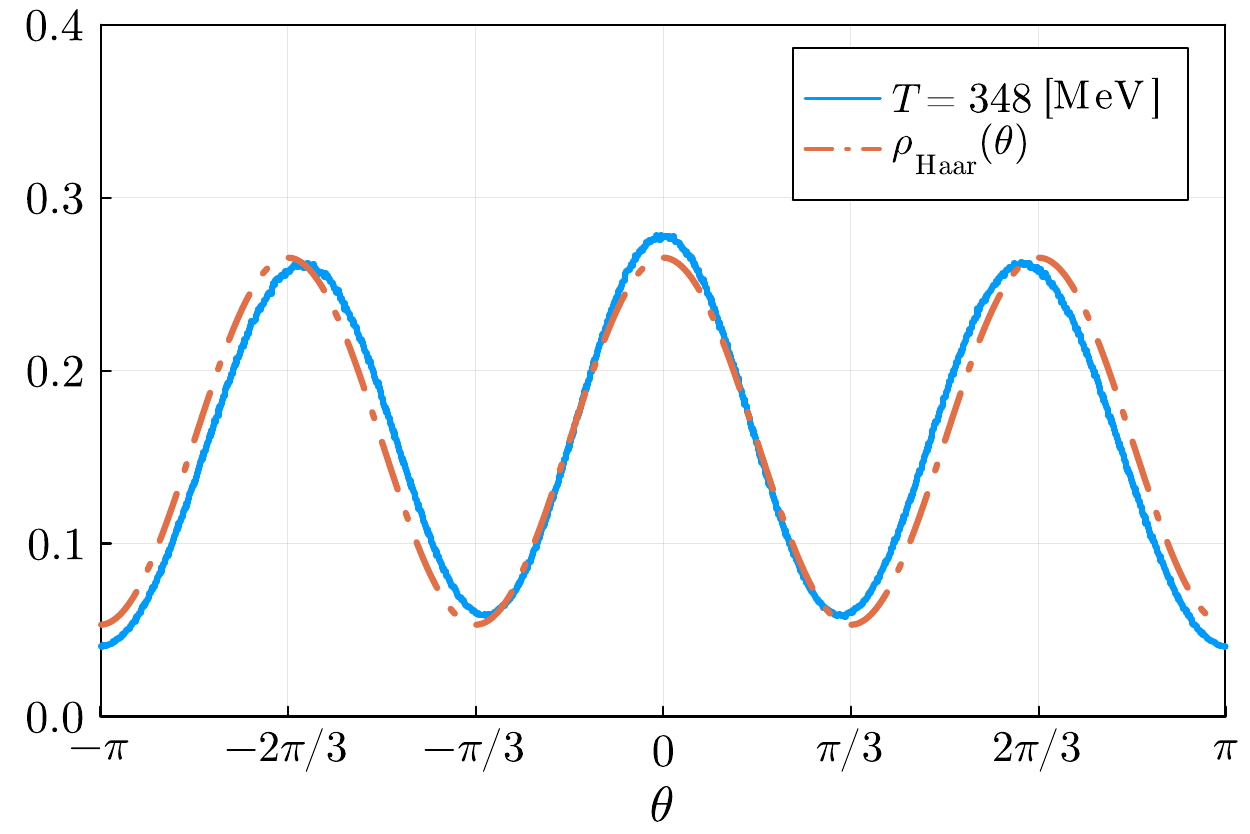}}
\scalebox{0.2}{
\includegraphics{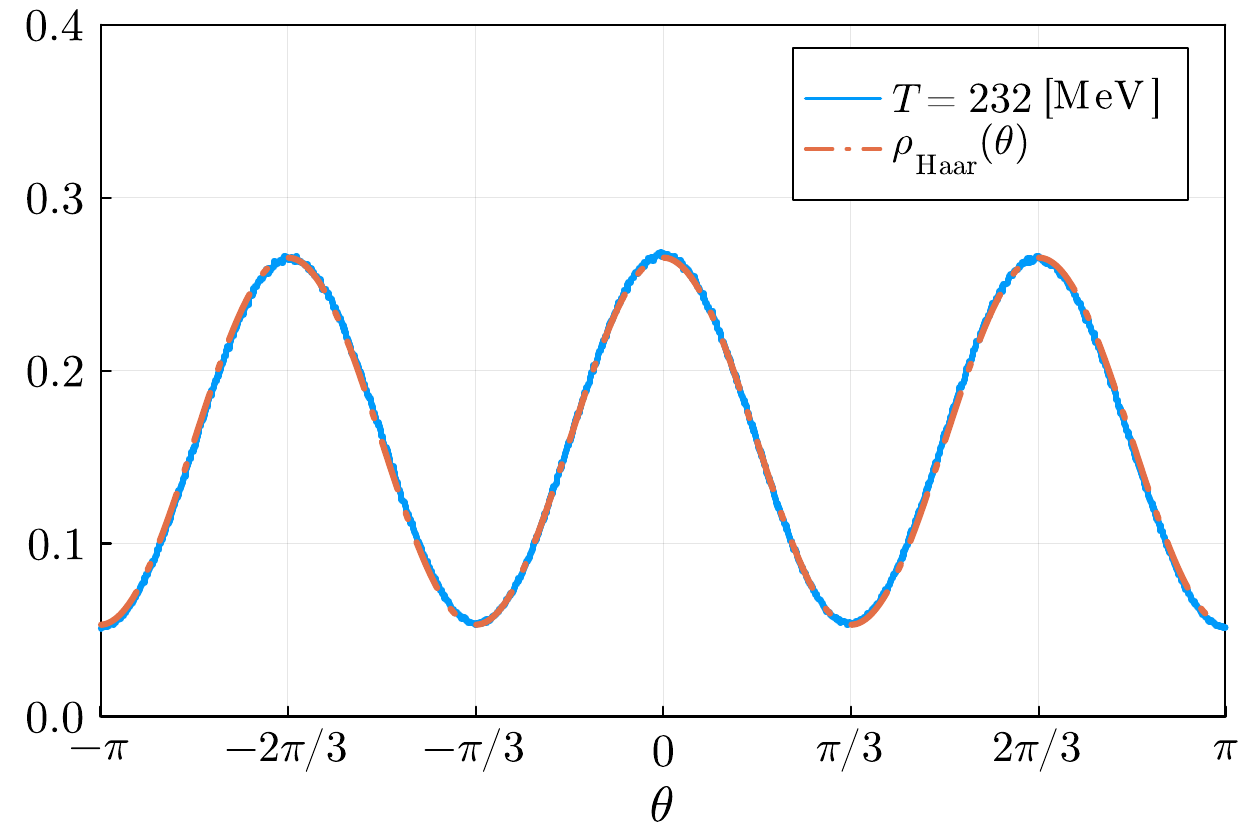}}
\end{center}
\caption{Distributions of Polyakov line phases obtained from the WHOT-QCD collaboration's configurations~\cite{Umeda:2012er}. Lattice size $N_t\times 32^3$  ($N_t=6,8,12$ and correspondingly $T =$ 464, 348, and 232 MeV.), 599 configurations for each temperature. 
Although the agreement with Haar-random distribution seems to be good at $T=232$ MeV, more careful investigation shows the small deviation and hence the onset of partial deconfinement; see Fig.~\ref{fig:T-vs-character}. 
}\label{fig:WHOT-phase}
\end{figure}

\begin{figure}[htbp]
\begin{center}
\scalebox{0.2}{
\includegraphics{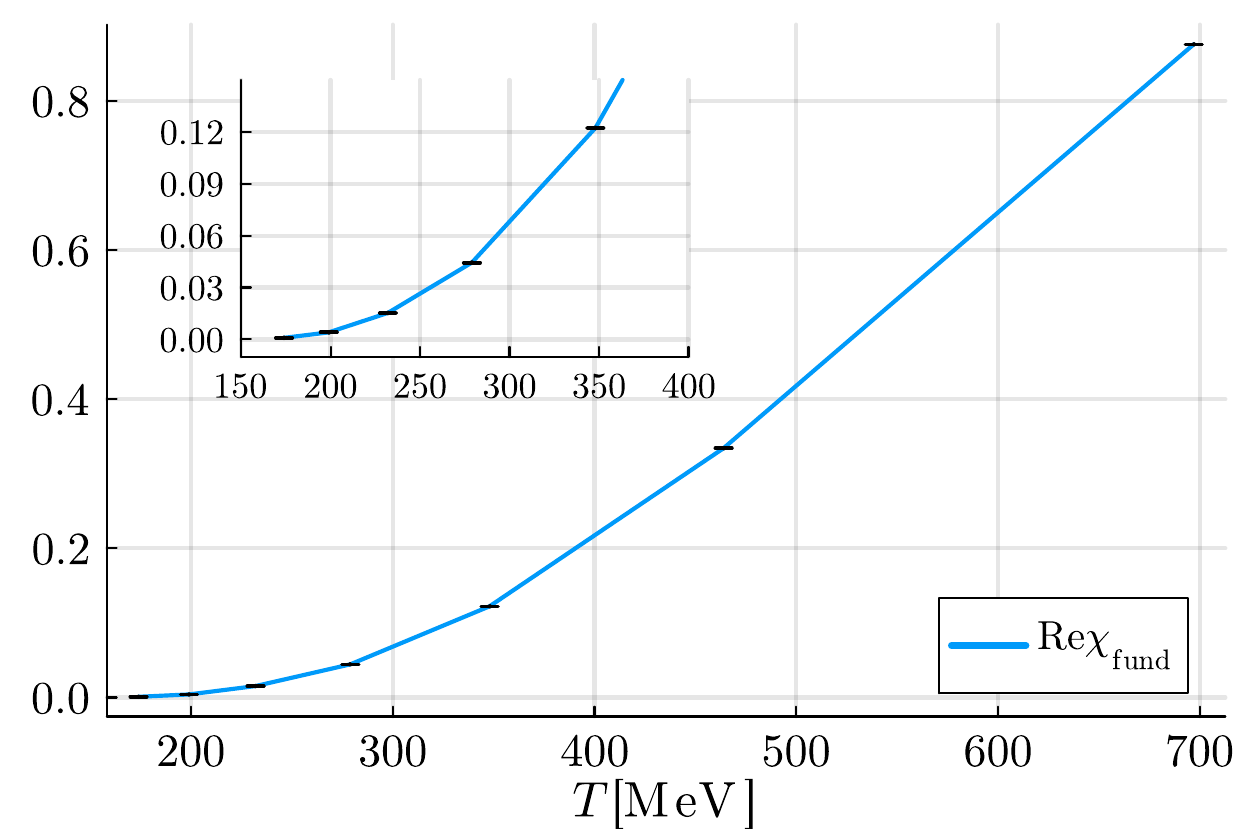}}
\scalebox{0.2}{
\includegraphics{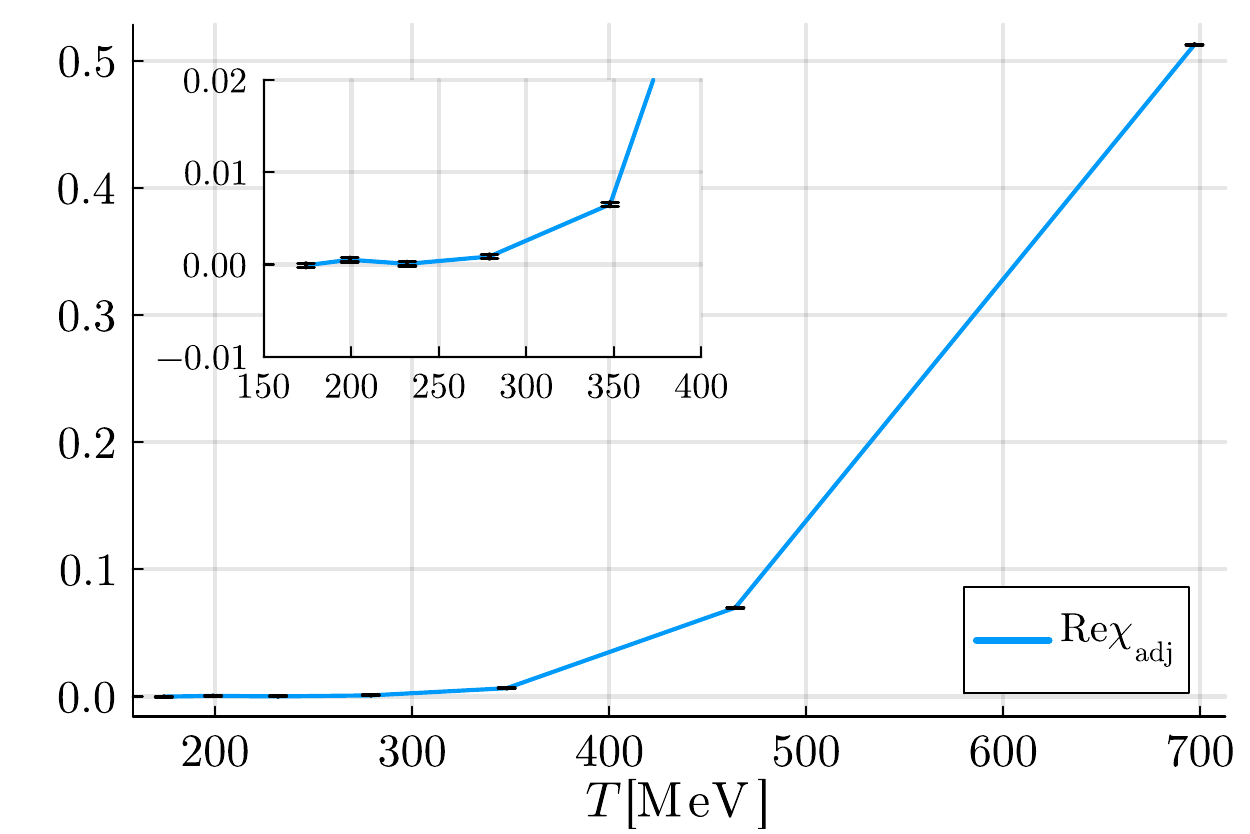}}
\scalebox{0.2}{
\includegraphics{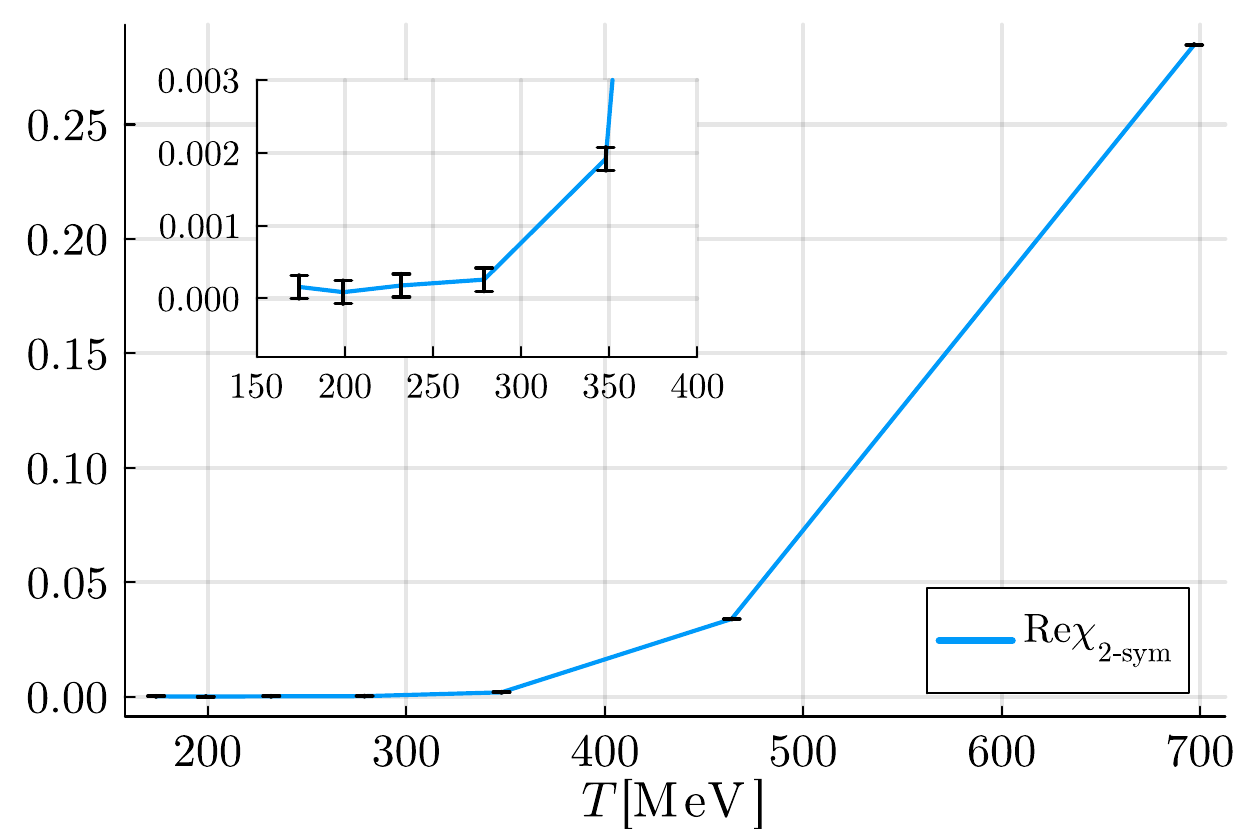}}
\end{center}
\caption{
The expectation values of characters vs. temperature for the fundamental, adjoint, and rank-2 symmetric representations, obtained from WHOT-QCD configurations. Note that the expectation values are real. 
}\label{fig:T-vs-character}
\end{figure}

In Fig.~\ref{fig:WHOT-phase}, we plot the distribution of Polyakov line phases obtained from the configurations created by the WHOT-QCD collaboration~\cite{Umeda:2012er}. 
We can see larger deviations from the Haar-random distribution at higher temperatures.  
To see the deviations quantitatively, we plot the expectation values of characters in Fig.~\ref{fig:T-vs-character}. The Haar-random distribution is equivalent to vanishing expectation value for all irreducible representations. We can see that the onset of the departure from Haar-random around the lowest temperature in our configuration set, $T=174$ MeV. This implies that  $T=174$ MeV is the completely-confined or partially-deconfined phase, and $T\ge 199$ MeV should be partially deconfined.

Can we identify the counterpart of the GWW transition at finite $N$? 
The GWW transition at large $N$ is associated with the condensation of Polyakov loops in large representations.
For SU(3) QCD, we can see that the fundamental Polyakov loop starts to increase at $T\sim 174$MeV while adjoint, rank-2 symmetric and rank-3 symmetric representations grow at $T\gtrsim 300$MeV. 
This observation suggest that complete deconfinement sets in at $T\sim300$MeV. 

When quarks are massless, the matching of 't Hooft anomaly requires the breaking of chiral symmetry in the confined phase. Combining it with partial deconfinement, we expect the chiral symmetry breaking below the GWW point, i.e., quarks in the confined sector should contribute to nonzero chiral condensate~\cite{Hanada:2019kue}. Given the lack of another characteristic energy scale, it is natural to expect that chiral symmetry breaking takes place at or near the GWW point. Numerical study of large-$N$ strong-coupling lattice gauge theory supports this conjecture~\cite{Hanada:2021ksu}. Chiral symmetry is explicitly broken when quarks are massive. Still, it would be natural to expect nontrivial signals at the GWW point. Our hypothesis is that instanton condenses below the GWW point. Hence we can pose a nontrivial conjecture: the condensation of Polyakov loops in large representations and instanton condensation should give us the same transition temperature.

\begin{figure}[hbtp]
\begin{center}
\scalebox{0.33}{
\includegraphics{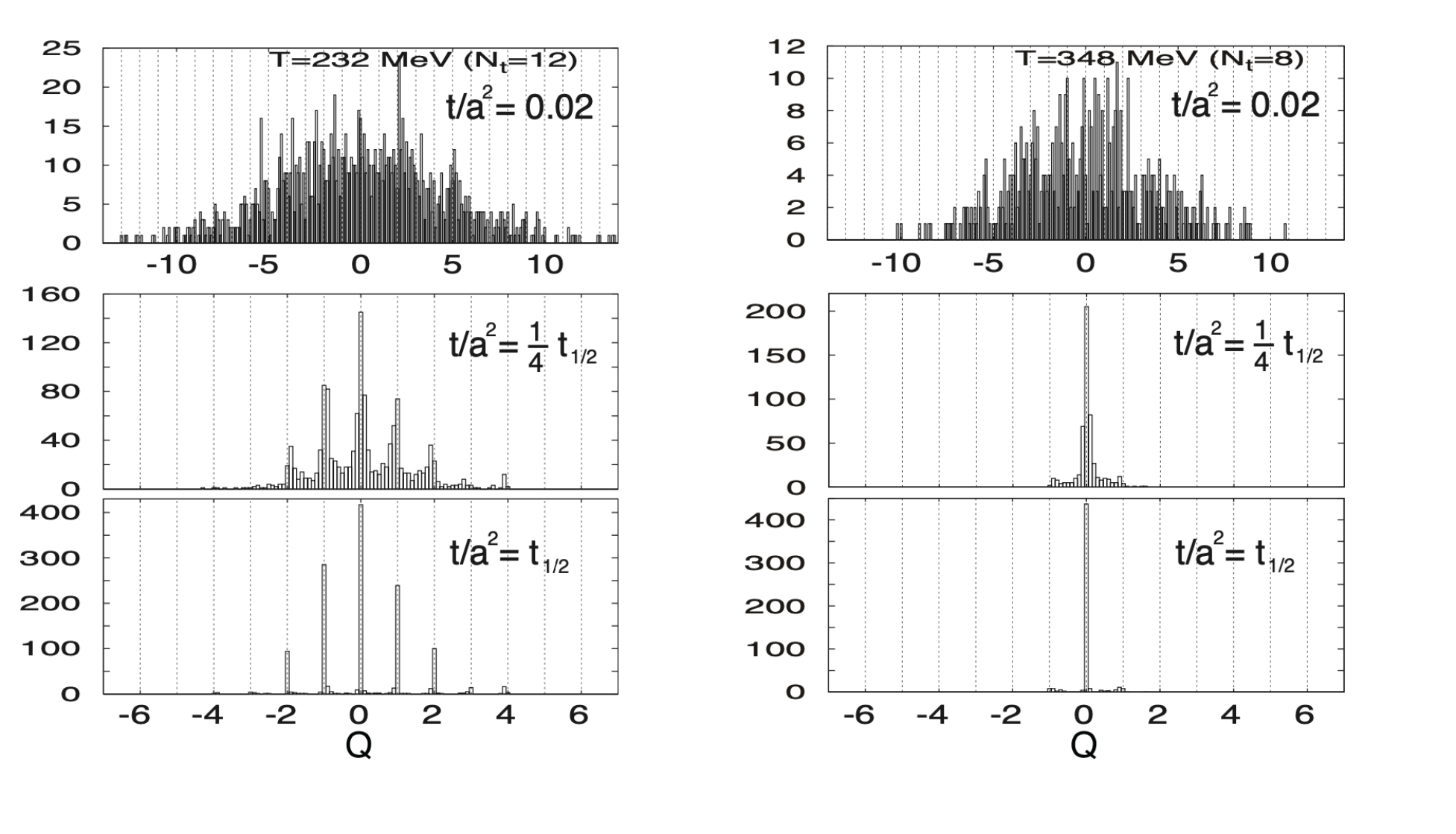}}
\end{center}
\caption{Histogram of the topological charge taken from Ref.~\cite{Taniguchi:2016tjc}. 
The parameter $t$ is the flow time. After sufficient smearing, charges peak at integer values.  At $T=232$ MeV, peaks at nonzero values signal the condensation of instantons. Flow time $t_{1/2}$ is defined in Ref.~\cite{Taniguchi:2016tjc} and chosen in such a way that unphysical effects from too much smearing can be avoided. For other temperatures, see Ref.~\cite{Hanada:2023rlk}. 
}\label{fig:topological-charge-WHOT}
\end{figure}

 To detect the instanton condensation, we use the topological charge computed by the WHOT-QCD collaboration~\cite{Taniguchi:2016tjc}.\footnote{
 The analysis of instanton condensation was completed after the presentation and reported in Refs.~\cite{Hanada:2023krw,Hanada:2023rlk}. 
 } To remove the UV-sensitivity of lattice configurations, they are smeared with the gradient flow~\cite{Luscher:2010iy}. 
 After smearing, the histogram of the topological charge has peaks at integer values as shown in Fig.~\ref{fig:topological-charge-WHOT}.
At $T\le 279$ MeV, multiple peaks are observed, signaling the instanton condensation. 
We can see wider distributions at lower temperatures. Peaks at $Q\neq 0$ disappear $T\gtrsim 348$ MeV. Therefore, we estimate that $T\lesssim$ $348$ MeV is the partially- or completely-confined phase. In this way, two different observables --- Polyakov loops in higher representations and topological charge -- give us the same estimate for the transition temperature. 

%%%%%%%%%%%%%%%%%%%
%%%%%%%%%%%%%%%%%%%
\section{Conclusion and discussion}\label{sec:conclusion}
%%%%%%%%%%%%%%%%%%%
%%%%%%%%%%%%%%%%%%%
We proposed finite-$N$ counterpart of partial deconfinement and provided some evidence based on lattice QCD data.
We claimed there are three phases (completely confined, partially deconfined and completely deconfined), although more analyses are needed in order to see these phases are separated by non-analyticity in partition function. 

On theoretical side, we used only universal properties applicable to any confining gauge theory. Therefore, we expect similar phase structure for other theories, and it is interesting to confirm such a phase structure by lattice simulations. Given the simplicity and universality of the underlying mechanism, we could expect some other universal features of confining gauge theories are explained based on the same idea. As an example, we note that the Casimir scaling of string tension is an almost immediate consequence of slowly varying Haar randomness~\cite{Bergner:2023rpw}.

%%%%%%%%%%%%%%%%%%%
%%%%%%%%%%%%%%%%%%%
\section*{Acknowledgement}
%%%%%%%%%%%%%%%%%%%
%%%%%%%%%%%%%%%%%%%
We would like to thank the members of the WHOT-QCD collaboration, including S.~Ejiri, K.~Kanaya, M.~Kitazawa, and T.~Umeda, for providing us with their lattice configurations and many plots and having stimulating discussions with us. 
The analysis of topological charge in Ref.~\cite{Taniguchi:2016tjc}, which was crucial in Sec.~\ref{sec:finite-N}, was led by Y.~Taniguchi who passed away in 2022. K.~Kanaya collected the data and plots created by Y.~Taniguchi for us. We deeply thank Y.~Taniguchi and K.~Kanaya. We also thank S.~Aoki, G.~Bergner, H.~Fukaya, V.~Gautam, Y.~Hayashi, J.~Holden, S.~Kim, T.~Kovacs, A.~Nakamura, M.~Panero, R.~Pisarski, E.~Rinaldi, Y.~Tanizaki, and J.~Verbaarschot for discussions and comments. This work was supported by the Japan Lattice Data Grid (JLDG) constructed over the SINET5 of NII. M.~H. thanks his STFC consolidated grant ST/X000656/1. H. O. was supported by a Grant-in-Aid for
JSPS Fellows (Grant No.22KJ1662). H. S. and H.W. were supported by the Japan Society for the Promotion
of Science (JSPS) KAKENHI Grant number 21H05182 and 22H01218, respectively.

\end{document}